\documentclass[conference, onecolumn]{IEEEtran}
\ifCLASSINFOpdf
   \usepackage[pdftex]{graphicx}
\else
\fi
\usepackage[cmex10]{amsmath}
\usepackage{bm}
\usepackage{hhline}
\usepackage{cite}
\usepackage{ifthen}
\usepackage{multirow}
\usepackage{csquotes}
\usepackage{caption}
\usepackage{subcaption}
\usepackage{setspace}
\usepackage{amssymb}
\usepackage{cleveref}
\usepackage{latexsym}
\usepackage{mathrsfs}
\usepackage{caption}
\usepackage{makecell}
\usepackage{color, soul}
\usepackage[switch]{lineno}

\usepackage{array}
\usepackage{tabulary}
\newcolumntype{K}[1]{>{\raggedright\arraybackslash}p{#1}}

\newcolumntype{C}[1]{>{\centering\let\newline\\\arraybackslash\hspace{0pt}}m{#1}}

\newcommand{\R}{\mathbb{R}}

\begin{document}
%
\title{ Generative Adversarial Networks and Conditional Random Fields for Hyperspectral Image Classification}
%
%
%


\author{Zilong Zhong$^{1,2}$, Jonathan Li$^{1,2,3}$, David~A.~Clausi$^{1,2}$, , Alexander Wong$^{1,2}$\\
$^1$ University of Waterloo, Waterloo, ON, Canada\\
$^2$Waterloo Artificial Intelligence Institute, Waterloo, ON, Canada\\
$^3$ Xiamen University, Xiamen, China\\}

\maketitle

\begin{abstract}
\normalsize
In this paper, we address the hyperspectral image (HSI) classification task with a generative adversarial network and conditional random field (GAN-CRF) -based framework, which integrates a semi-supervised deep learning and a probabilistic graphical model, and make three contributions. First, we design four types of convolutional and transposed convolutional layers that consider the characteristics of HSIs to help with extracting discriminative features from limited numbers of labeled HSI samples. Second, we construct semi-supervised GANs to alleviate the shortage of training samples by adding labels to them and implicitly reconstructing real HSI data distribution through adversarial training. Third, we build dense conditional random fields (CRFs) on top of the random variables that are initialized to the softmax predictions of the trained GANs and are conditioned on HSIs to refine classification maps. This semi-supervised framework leverages the merits of discriminative and generative models through a game-theoretical approach. Moreover, even though we used very small numbers of labeled training HSI samples from the two most challenging and extensively studied datasets, the experimental results demonstrated that spectral-spatial GAN-CRF (SS-GAN-CRF) models achieved top-ranking accuracy for semi-supervised HSI classification. 
\end{abstract}


%
\IEEEpeerreviewmaketitle

\section{Introduction}
%
%
%
%

Due to their hundreds of spectral bands, the accurate interpretation of hyperspectral images (HSIs) has attracted significant scholarly attention from the machine learning and remote sensing communities \cite{li2014hyperspectral,jia20183,yuan2016hyperspectral,zhou2016learning}. Recent studies suggest that supervised deep learning models can alleviate challenges caused by the high spectral dimensionality of HSIs and achieve strikingly better classification accuracy \cite{Chen2016Deep,zhong2017spectral,Luo2018}. However, there are still three challenges that prevent deep learning models from offering precise pixel-wise HSI classification maps \cite{tarabalka2010segmentation,yuan2017spectral}. First, the high dimensionality of HSI pixels make it hard to directly use the deep learning models for normal optical images in HSI interpretation. Second, the shortage of labeled pixels limits the classification performance of deep learning models. Third, the classification maps generated by deep learning models tend to be noisy and have spurious object edges. In this proposal, I will analyze these challenges and offer our suggestions to mitigate them. 

The first challenge derives from the high spectral dimensionality of HSIs, which comprise two spatial dimensions and one spectral dimension. Conventional methods focus on reducing the spectral dimensionality of HSIs. For example, \cite{Zhao2016Spec} adopted dimensionality reduction methods to extract discriminative features for the convolutional neural networks (CNNs) that follow to classify. Nonetheless, these methods ignore the inherent dimension reduction capability of deep learning models. Many papers indicate that both spectral and spatial features play important roles in precise HSI interpretation. For instance, \cite{yuan2017spectral} employed a shared structured learning strategy to construct a discriminant linear projection for spectral-spatial HSI classification. Additionally, \cite{Chen2016Deep} proposed an end-to-end CNN model for HSI classification and achieved promising results, which show the generality of deep learning models. However, most machine learning models for HSI interpretation overlook the characteristics of this remotely sensed data.

The second challenge stems from the high cost of and difficulty in obtaining a large amount of labeled data for HSIs.  Deep learning models \cite{Li2017Hyperspectral,Chen2014Deep,Chen2015Spec} are prevailing for HSI classification. Many papers suggest that these models require a large amount of training data. For example, \cite{Chen2016Deep} provided a method adding noise to HSI pixels in order to increase the number of training samples. Additionally, \cite{Li2017Hyperspectral} proposed a pixel-pair approach that samples two pixels independently and couples them as a group for the purpose of enlarging the training data size. The scarcity of training samples is especially the case for some land cover classes in HSI data sets. However, in contrast to the conventional optical image classification objectives in the computer vision domain \cite{Krizhevsky2012Imagenet,LeCun2015Deep}, which usually contain hundreds or thousands of classes, the land cover classification objectives of HSIs have far fewer classes to recognize. Therefore, the assumption that deep learning models require a high amount of data for training may not hold for HSIs. On the other hand, a large amount of unlabeled data remains an unexploited gold mine for efficient data use. Several works that focus on semi-supervised learning used small numbers of labeled and large numbers of unlabeled HSI samples for training. For instance, \cite{Mnih2015Human} adopted multi-layer neural networks to propagate labels from annotated HSI pixels to unannotated ones. Moreover, \cite{Chen2014Deep} applied a stacked convolutional autoencoder to use spectral-spatial representation learned from other HSI datasets. Although this paper achieved accurate classification results, these results may originate from the large area of spatial information contained in each training sample rather than deep learning models.

The third challenge is caused by the complexity of HSIs. Multiple works utilize the smoothness assumption that favors geometrically simple classification results \cite{cao2018hyperspectral,Tarabalka2010SVM,zhong2014hybrid,yang2014semi}. For example, \cite{Tarabalka2010SVM} incorporated a probabilistic graphical model as the post-processing step to improve the classification outcomes of kernel support vector machines (SVMs). \cite{zhong2011modeling} constructed a conditional random field (CRF) with a high-order term to consider more complex relationships between different spectral bands and obtained very promising outcomes. Additionally, \cite{zhong2014hybrid} incorporated a CRF for pre-processing as well as post-processing to stress the a priori smoothness and refine the classification maps. The adoption of probabilistic graphical models on top of supervised classification models can also be conceived as a way to take the unlabeled samples into account for HSI classification because this step does not require the ground truth annotation of neighboring pixels. However, most CRF based models consider only the short-range correlations of pixels and ignore the long-range ones.

\begin{figure*}[!t]
\centering
\includegraphics[width=7in]{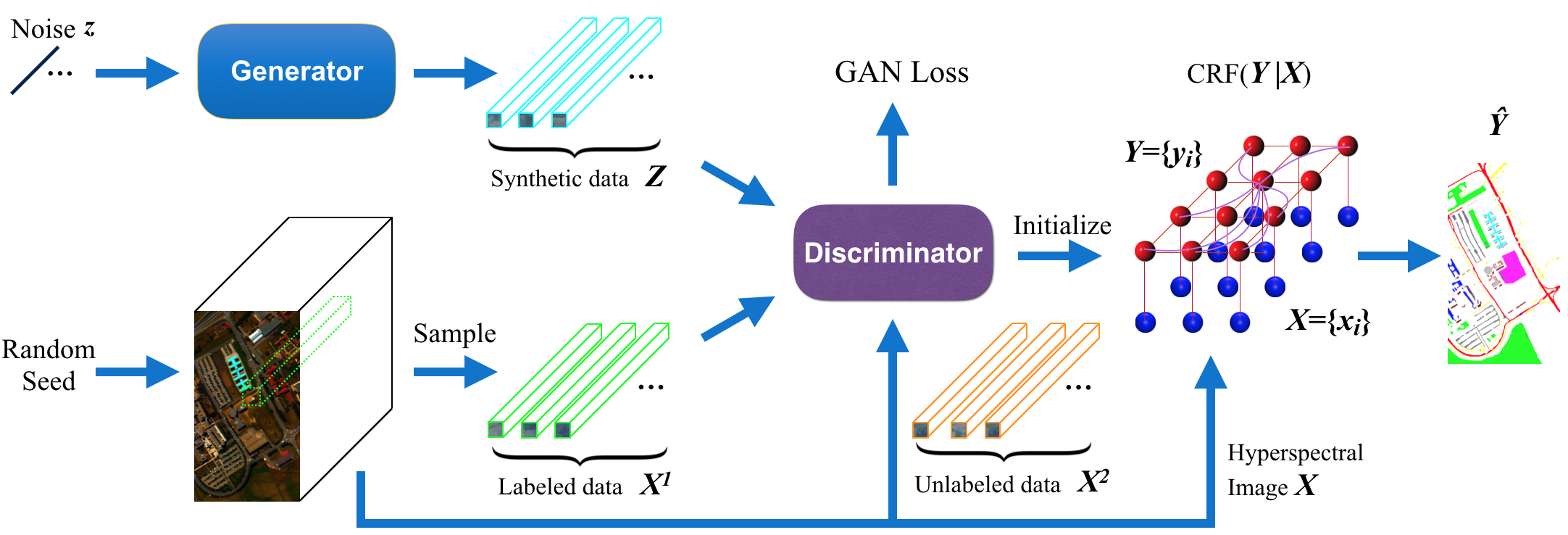}%
\caption{A semi-supervised GAN-CRF framework for HSI classification. First, in the semi-supervised GAN, a generator transforms noise vectors $\boldsymbol{z}$ to a set of fake HSI cuboids $\boldsymbol{Z}$, and a discriminator tries to distinguish the categorical information as well as the genuineness of input cuboids that come from $\boldsymbol{X}^1$ or $\boldsymbol{Z}$. Then, a dense CRF is established by using the softmax prediction of the trained discriminator about $\boldsymbol{X}^2$ to initialize random variables $\boldsymbol{Y}$, which is conditioned on the HSI data $\boldsymbol{X}$. Mean field approximation is adopted to offer a refined classification map $\boldsymbol{\hat{Y}}$ for the post-processing CRF.}
\label{fig_1}
\end{figure*}

In the face of these difficulties, two common semi-supervised learning methods --- graph-based models and generative models --- have been adopted to alleviate them \cite{yang2014semi,ji2014spectral, zhan2018semisupervised,zhu2018generative}. Graph-based models are premised on the smoothness assumption that accentuates geometrically simple classification results. For example, \cite{yang2014semi} imposed a manifold regularizer on a Laplacian SVM framework to learn spectral-spatial features for HSI image classification. Additionally, \cite{ji2014spectral} proposed a dual hypergraph framework that imposes spectral-spatial constraints by jointly calculating a Laplacian matrix. Although these graph-based semi-supervised methods take both labeled and unlabeled samples into account, they identify HSI pixels based on hand-crafted features. Generally, these features learned from feature engineering steps are difficult to tune or generalize to other cases. Moreover, the performance of these semi-supervised models largely depends on the quality of unlabeled data, which is hard to control or standardize. Recently, a generative model called generative adversarial network (GAN)\cite{Goodfellow} has attracted a lot of attention for image generation. For instance, \cite{zhan2018semisupervised} proposed a semi-supervised 1D-GAN for HSI classification, but ignored the spatial attribute of HSIs that can be used for enhancing classification performance. Moreover, \cite{zhu2018generative} used convolutional neural networks (CNNs) to build generative adversarial networks for HSI classification and achieved very promising results. However, the discriminators used in this paper only use three principled component analysis (PCA) channels of HSIs and therefore do not fully exploit the spectral characteristic of HSIs.  

\begin{figure*}[!t]
\centering
\includegraphics[width=7.1in]{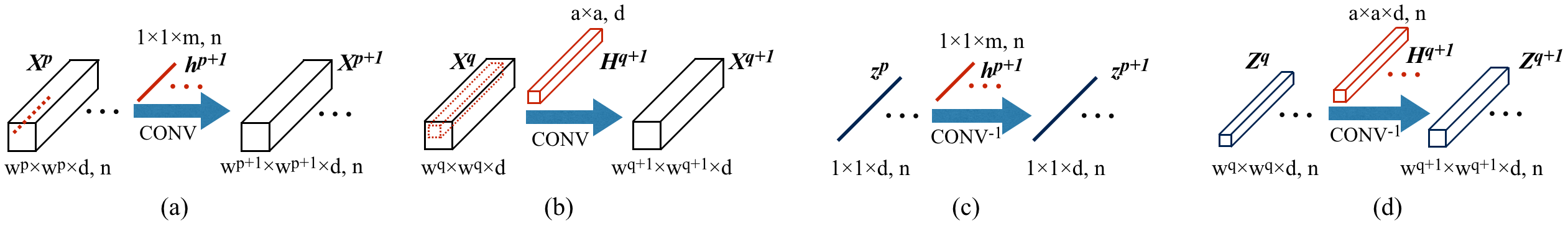}%
\caption{Four basic convolutional and transposed convolutional  layers aiming for hyperspectral features extraction and generation in semi-supervised GAN-CRF models. (a) - (b) Spectral and spatial convolutional layers in discriminators. (c) - (d) spectral and spatial transposed convolutional layers in generators.}
\label{fig_2}
\end{figure*}

Inspired by \cite{Goodfellow} and \cite{Chen_Deeplab}, we suggest a semi-supervised deep learning framework that consists of a generator, discriminator, and conditional random field built on top of the discriminator. The discriminator and generator form a generative adversarial network based on game theory. Specifically, the discriminator adopts spectral-spatial convolutional layers to learn discriminative features from a small amount of labeled data and unlabeled data, and the generator employs spectral-spatial transposed convolutional layers to reconstruct HSI samples from vectors of Gaussian noise. Unlike traditional semi-supervised models, which require a large amount of unlabeled data for training, our proposed framework is data-efficient because the generator creates a high amount of synthetic data and the discriminator takes a small number of unlabeled samples. In this way, the GAN-CRF model estimates the real data distribution, mitigates the shortage of annotated data, and smooths the semi-supervised learning process. In addition, the output of the discriminator is the unary input term of the subsequent CRF. The binary term of the CRF imposes an a priori smoothness whereby adjacent pixels are more likely to belong to the same categories. More importantly, the CRF takes on a fully connected form that imposes a random field on the whole classification map and considers the long-range relationship between HSI pixels. Thus, by taking a generative adversarial network and considering the continuity of neighboring pixels, the designed semi-supervised architectures learn local fine-grain representation as well as high-level invariant features of HSI pixels concurrently.

The main contributions of this study are as follows:
\begin{enumerate} 

\item   We integrate the spectral-spatial attributes of HSIs into convolutional and transposed convolutional layers of a GAN-CRF framework to learn discriminative spectral-spatial features of HSI samples.

\item	We construct semi-supervised GANs to alleviate the shortage of labeled data through adversarial training, which is a zero-sum game between the discriminators and generators of GANs.

\item	We build dense conditional random fields that impose graph constraints on the softmax predictions of trained discriminators to refine HSI classification maps. 

\end{enumerate}

The overall structure of this paper takes the form of five sections: Section II reviews related works with regard to the GAN-CRF framework. Section III introduces fundamental layers, spectral-spatial discriminators and generators, semi-supervised GANs, and post-processing CRFs of GAN-CRF models. Section IV offers model parameter settings, comparative experiments, and discussions. Last, section V makes some conclusions.

\section{Related Work}

GANs are unsupervised deep learning models that provide a solution to implicitly estimate real data distribution and correspondingly generate synthetic samples. Recently, there has been increasing interest in GANs for unsupervised learning, especially in regards to generating synthetic images that approximate the distribution of real ones \cite{Goodfellow,Salimans}. Compared with traditional generative methods, GANs are not constrained by Markov fields or explicit approximation inference. For instance, a deep convolutional GAN \cite{Radford} that consists of deep convolutional layers has been proposed to generate high-quality images. The original GAN aims for image generation and its variants have generated astonishing controllable and partially explainable images \cite{saatci2017bayesian}. The GAN employs a discriminator and a generator to compete with each other. Specifically, the generator generates synthetic examples to deceive the discriminator, and the discriminator distinguishes real samples from fake ones. Since their objectives are contradictory, the training of the discriminator and generator of a GAN can be regarded as a process to find a Nash equilibrium through a game-theoretical point of view. Therefore, this GAN training can be formulated as a min-max optimization problem:

\begin{equation}
\begin{split}
\min_G\max_D Loss(D,G) &= E_{\boldsymbol{x}\sim p_{data}}[\log{D(\boldsymbol{x})}] + E_{\boldsymbol{z}\sim p_{z}}[\log{(1-D(G(\boldsymbol{z})))]},
\end{split}
\end{equation}

where $D(\cdot)$ and $G(\cdot)$ represent   softmax outputs of a discriminator and synthetic data generated by a generator, respectively. $\boldmath{x}$ and $\boldmath{z}$ denote true images and vectors of Gaussian noise, and they follow the distributions of real HSI data and Gaussian noise, respectively.
GANs produce  very promising image generation results in datasets like the MNIST digit database \cite{lecun1998gradient} and the Yale Face database \cite{yang2004two}, both of which contain compact data distribution and similar image layout. 

Graph models have widely been used for remotely sensed image interpretation tasks to effectively impose smoothness constraints on classification or segmentation results\cite{zhao2016high, Chen_Deeplab}. CRFs are graphical models that assume a priori continuity whereby neighboring pixels of similar spectral signatures tend to have the same labels \cite{zhong2011modeling}. Since CRFs can be regarded as a structured generalization of multinomial logistic regression, the conditional probability distribution of a CRF takes this form:

\begin{equation}
Prob(\boldsymbol{y}|\boldsymbol{x}) = \frac{\exp(-E(\boldsymbol{y}|\boldsymbol{x}))}{\sum_{\boldsymbol{y}}\exp(-E(\boldsymbol{y}|\boldsymbol{x}))},
\label{crf}
\end{equation}       
where $\boldsymbol{y}$ and $\boldsymbol{x}$ denote output random variables and their corresponding observed data. $E(\cdot)$ is an energy function that models the joint probability distribution of $\boldsymbol{y}$ and $\boldsymbol{x}$. The optimal random variables can be calculated by the \textit{maximum a posteriori} (MAP) estimation:
\begin{equation}
\boldsymbol{y}^{MAP} =\underset{\boldsymbol{y}}{argmax} \,Prob(\boldsymbol{y}|\boldsymbol{x}).
\end{equation}
However, although Equation \eqref{graph} usually is an intractable problem, it can be solved through approximation methods \cite{krahenbuhl2011efficient}. 

\section{Proposed Model}

\begin{figure*}[!t]
\centering
\includegraphics[width=7in]{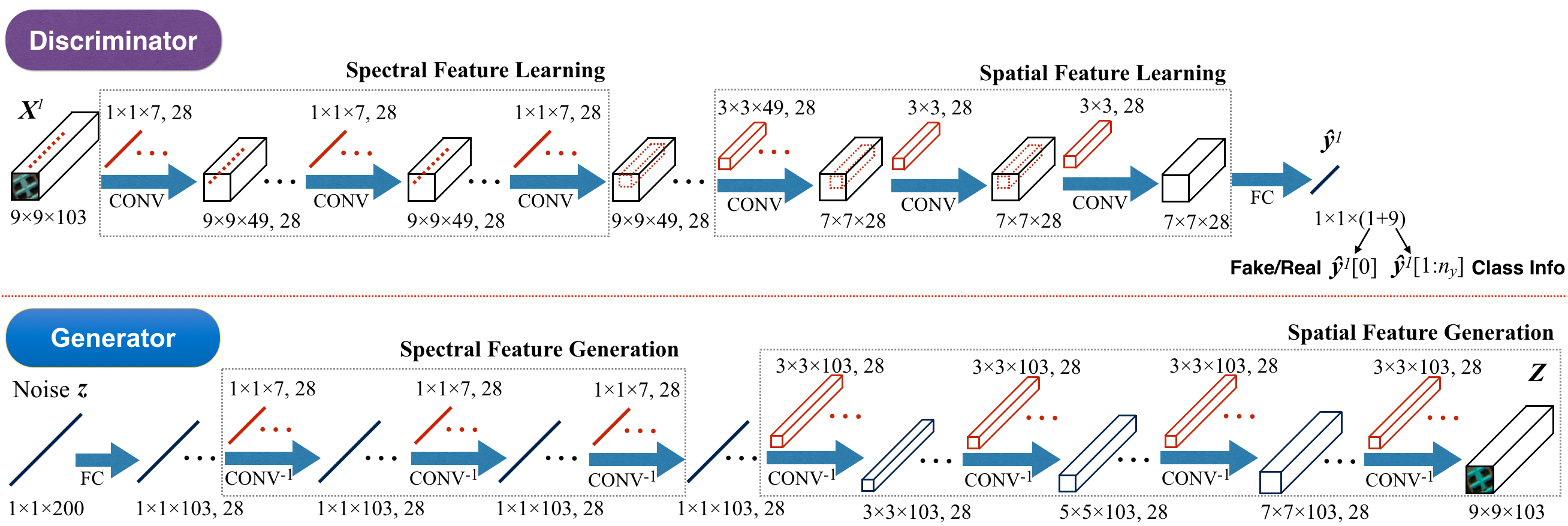}
\caption{A spectral-spatial discriminator (upper), which comprises consecutive spectral and spatial feature learning blocks, outputs a vector that contains a indicative entry of fake or real and categorical probabilities; and a spectral-spatial generator (lower), which comprises consecutive spectral and spatial feature generation blocks, transforms a vector of Gaussian noise to a synthetic HSI cuboid.}
\label{fig_3}
\end{figure*}

To solve the three challenges of HSI classification, we propose a GAN-CRF -based semi-supervised deep learning framework. Suppose a hyperspectral image $\boldsymbol{X}$ contains $m$ pixels 
$\{ \boldsymbol{x_i}\} \in \R^{ n_x \times m }$, where $n_x$ represents the number of spectral bands. Then, we sample two groups of HSI cuboids from $\mbox{\boldmath{$X$}}$: the labeled group $\boldsymbol{X}^1 = \{ \boldsymbol{X_i}^1 \} \in \R^{ n_x \times w \times w \times m_l}$ and the unlabeled group $\boldsymbol{X}^2 = \{ \boldsymbol{X_i}^2 \} \in \R^{ n_x \times w \times w \times m_u}$, where $w$, $m_l$, and $m_u$ are the spatial width of HSI cuboids, the number of labeled, and the number of unlabeled HSI samples, respectively. Since each pixel in $\boldsymbol{X}$ corresponds to a HSI cuboid in $\{ \boldsymbol{X_i}^1, \boldsymbol{X_i}^2 \}$, therefore $m=m_l+m_u$. The labeled group $\boldsymbol{X}^1$ has its annotation $\boldsymbol{Y}^1 = \{ \boldsymbol{y_i}^1 \} \in \R^{ (1+n_y)  \times m_l}$, where $n_y$ is the number of land cover classes and $\boldsymbol{y_i}^1[0]$ (the first entry in a vector $\boldsymbol{y_i}^1$) indicates whether the corresponding HSI cuboid is fake (1/0 means fake/real). As shown in Figure \ref{fig_1}, the whole model is composed of a discriminator, a generator, and a post-processing CRF. Since annotations $\boldsymbol{Y}^1$ of real HSI samples are used for training, the discriminator and generator form a semi-supervised GAN. The generator transforms noise vectors $\mbox{\boldmath{$z$}}$ to synthetic HSI cuboids $\boldsymbol{Z} = \{\boldsymbol{Z_i}\}$, each sample of which have the same size as those from $\boldsymbol{X}^2$. The discriminator attempts to distinguish real HSI cuboids $\boldsymbol{X}^1$ from fake ones $\boldsymbol{Z}$ and to classify real HSI cuboids.

In contrast to updating one discriminative model in supervised deep learning, the training of a GAN involves searching an equilibrium between the generator and discriminator by using stochastic gradient descent or similar methods to optimize the parameters of the GAN. However, GANs are known for their instability in training, and it is almost impossible to find an optimal equilibrium between their generators and discriminators. Therefore, we adopt an alternating optimization strategy that successively updates the parameters of the generator and discriminator in each training iteration to help the discriminator to learn discriminative features using a small amount of labeled data and a large amount of synthetic data produced by the generator. When the training of a GAN is completed, we use the trained discriminator of the GAN to make a prediction about the unlabeled group  $\boldsymbol{X}^2$. Then, a conditional random field is established by using the softmax predictions of the trained discriminator to initialize random variables $\boldsymbol{Y}=\{\boldsymbol{y_i}\} \in \R^{ (1+n_y)  \times m}$ that are conditioned on the raw HSI $\boldsymbol{X}$. Last, we use mean field approximation to optimize the conditional random field and get a refined classification map $\boldsymbol{\hat{Y}}$.

\subsection{Spectral-Spatial Discriminator and Generator}

Discriminative deep learning models, such as CNNs and their extensions, have been used for HSI feature extraction and they have substantially outperformed traditional machine learning methods given enough training data\cite{Chen2016Deep,zhong2017spectral}. However, both these approaches ignore the inherent difference in spectral dimensionality between hyperspectral images and common images used in computer vision tasks. Based on the assumption that the sampled HSI data form a low dimensional manifold embedded in a higher dimensional space, multiple models have tried to reduce the high dimensionality of HSI pixels and to learn more efficient representation \cite{Zhang2014Hyperspectral,Zhao2016Spec}. However, the dimension reduction process inevitably leads to the loss of useful information. 

The specialty of HSI samples lies in its high spectral dimensionality.  
Recently, in response to this characteristic, \cite{zhong2017spectral} implemented a spectral-spatial residual network (SSRN) that considers the characteristics of HSI by consecutively extracting spectral and spatial features and obtained state-of-the-art supervised classification results. Therefore, as illustrated in Figure \ref{fig_2} (a)-(b), we extend the idea of spectral and spatial convolution from  \cite{zhong2017spectral} to the discriminator of a GAN-CRF model. If $\boldsymbol{X}^{[p+1]}$ and $\boldsymbol{X}^{[q+1]}$ represent the feature tensors of $[p+1]$th spectral and $[q+1]$th spatial convolutional layers, then the spectral and spatial convolutional layers of a discriminator can be formulated as follows: 

\begin{equation}
\boldsymbol{X}^{[p+1]} = LReLU( \boldsymbol{w}^{[p+1]} * \boldsymbol{X}^{[p]} + b^{[p+1]}),
\end{equation}

\begin{equation}
\boldsymbol{X}^{[q+1]} = LReLU(\boldsymbol{W}^{[q+1]} * \boldsymbol{X}^{[q]}+ b^{[q+1]}),
\end{equation}
where $\boldsymbol{w}^{[p+1]}$ and $\boldsymbol{W}^{[q+1]}$ represent the  $[p+1]$th spectral and $[q+1]$th spatial  convolutional kernels, respectively. $b^{[p+1]}$ and $b^{[q+1]}$ are the biases of these two layers. $*$ denotes the convolutional operation. $LReLU(\cdot)$ is a leaky rectified linear unit function:

\begin{equation}
LReLU(a) = 
\begin{cases}
 	a,				& \text{if a $>$ 0}, \\
    0.2a,          & \text{otherwise}.
\end{cases}
\end{equation}

In this work, we use padding tricks to keep the spatial size of feature tensors in most convolutional layers unchanged. The goal of adopting spectral-spatial convolutional layers in a GAN-CRF model is to exploit as much information as possible from limited labeled HSI samples. Similarly, we stretch the spectral-spatial idea to transposed convolutional layers. As shown in Figure \ref{fig_2} (c)-(d), the spectral and spatial transposed convolutional layers of a generator can be formulated as follows:

\begin{equation}
\boldsymbol{z}^{[p+1]} = ReLU(\boldsymbol{h}^{[p+1]} *^T \boldsymbol{z}^{[p]}  + b^{[p+1]}),
\end{equation}

\begin{equation}
\boldsymbol{Z}^{[q+1]} = ReLU( \boldsymbol{H}^{[q+1]} *^T \boldsymbol{Z}^{[q]}  + b^{[q+1]}),
\end{equation}
where $\boldsymbol{h}^{[p+1]}$ and $\boldsymbol{H}^{[q+1]}$ represent the $[p+1]$th transposed spectral and $[q+1]$th transposed spatial convolutional kernels. $b^{[p+1]}$ and $b^{[q+1]}$ are the biases of these two layers. $*^T$ denotes the transposed convolutional operation. $ReLU(\cdot)$ is the rectified linear unit function:

\begin{equation}
ReLU(a) = 
\begin{cases}
 	a,			& \text{if a $>$ 0}, \\
    0,          & \text{otherwise}.
\end{cases}
\end{equation}

As shown in Figure \ref{fig_2}, in contrast to spatial convolutional layers, the transposed convolutional layers expand the spatial size of feature tensors. In both the discriminator and generator of a GAN-CRF model, we apply batch normalization \cite{Ioffe2015Batch} in all convolutional and transposed convolutional layers to stabilize the training of a GAN.

\subsection{Semi-supervised GAN}

A GAN can be regarded as a combination of discriminative and generative models, where the discriminator focuses on learning discriminative features, and the generator concentrates on implicitly reconstructing real data distribution from random noises. As an example of University of Pavia (UP) dataset shown in Figure \ref{fig_3}, the discriminator comprises three spectral convolutional layers, three spatial convolutional layers, and a fully connected layer before a vector of softmax outputs. Conversely, the generator consists of a fully connected layer, three transposed spectral convolutional layers, and four spatial transposed convolutional layers to produce a synthetic hyperspectral cuboid. 

As the generator of a GAN can produce reasonable synthetic images and utilize them to train the discriminator of the GAN, many research papers have extended the discriminator of GANs to semi-supervised classification \cite{saatci2017bayesian,dai2017good,zhan2018semisupervised}. Similarly, we generalize the GAN to the semi-supervised HSI classification task. Since the labeled hyperspectral cuboid group $\boldsymbol{X}^1 = \{\boldsymbol{X}_i^1\}$ has its corresponding annotation group $\boldsymbol{Y}^1 = \{\boldsymbol{y}_i^1\}$ , the prediction of trained discriminators take this form:

\begin{equation}
\label{disc}
\hat{\boldsymbol{Y}}^1 = D(\boldsymbol{X}^1;\theta_D),
\end{equation}
each element $\hat{\boldsymbol{y}}_i^1$ of which has $(1+n_y)$ entries. Specifically, $\hat{\boldsymbol{y}}_i^1[0]$ indicates the genuineness of a hyperspectral cuboid, and $\hat{\boldsymbol{y}}_i^1[1:n_y]$ is a vector of softmax outputs that shows the probabilities of a hyperspectral cuboid belonging to the $n_y$ land cover classes. Compared to the original GAN that discriminates real data from fake ones, a semi-supervised GAN recognizes the categorical information of HSI cuboids by adding a supervised term to the loss function of a GAN. 

It is worth noting that the objectives of an unsupervised GAN and a semi-supervised GAN are different and even partially contradictory. The unsupervised GAN aims for implicitly estimating the true data distribution. On the contrary, the semi-supervised GAN focuses on data generation using limited labeled samples. Therefore, training a  semi-supervised GAN jeopardize its image generation capability. As presented in \cite{dai2017good}, a good semi-supervised GAN requires a bad generator because this generator produces data outside real data distribution, which in turn helps the discriminator recognizes real data more accurately. In this way, the generator that produces synthetic HSI cuboids functions as a regularizer on the discriminator. Therefore, the loss function regarding optimize the discriminator of a GAN for semi-supervised HSI classification takes the form:

\begin{equation}
L_{SEMI} (\theta_D, \theta_G) = L_{SUP}(\theta_D) + L_{D1}(\theta_D) + L_{D2}(\theta_D, \theta_G),
\end{equation}
where $\theta_{D}$ and $\theta_{G}$ are the parameters of a discriminator and a generator, respectively. $L_{SEMI}$ is the total semi-supervised loss for training the discriminator of a semi-supervised GAN, $L_{SUP}$, $L_{D1}$, and $L_{D2}$ represent the supervised loss of a discriminator, the unsupervised loss of a discriminator, and the unsupervised loss of a generator, respectively. These three terms are formulated as follows:

\begin{equation}
\begin{split}
L_{SUP} (\theta_D) &= -E_{\boldsymbol{X}^1\sim p_{data}}\log{D(\boldsymbol{X}^1; \theta_D)[1:n_y]} \\
&= -E_{\boldsymbol{X}^1\sim p_{data}}\log{\hat{\boldsymbol{Y}}^1[1:n_y]},
\end{split}
\end{equation}

\begin{equation}
\begin{split}
L_{D1} (\theta_D) &= - E_{\boldsymbol{X}^1\sim p_{data}}\log({1-D(\boldsymbol{X}^1; \theta_D)[0])} \\
&= - E_{\boldsymbol{X}^1\sim p_{data}}\log({1-\boldsymbol{\hat{Y}^1}[0]}),
\end{split}
\end{equation}

\begin{equation}
\begin{split}
L_{D2} (\theta_D, \theta_G) 
&= - E_{\boldsymbol{z}\sim p_{z}} \log{D(G(\boldsymbol{z}; \theta_G); \theta_D)} [0] \\
&= - E_{\boldsymbol{z}\sim p_{z}} \log{D(\boldsymbol{Z}; \theta_D)} [0] \\
&= - E_{\boldsymbol{z}\sim p_{z}} \log \boldsymbol{\hat{Y}}^1 [0].
\end{split}
\end{equation}
It is worth mentioning that $L_{D1} + L_{D2}$ also is the part of the total semi-supervised loss $L_{SEMI}$ that aims at training the bad generator of a GAN \cite{dai2017good}. Correspondingly, the loss function for training the generator of a semi-supervised GAN takes this form:

\begin{equation}
\begin{split}
L_{G} (\theta_D, \theta_G) 
&= - E_{\boldsymbol{z}\sim p_{z}} \log{(1-D(G(\boldsymbol{z}; \theta_G); \theta_D)} [0]) \\
&= - E_{\boldsymbol{z}\sim p_{z}} \log{(1-D(\boldsymbol{Z}; \theta_D)} [0]) \\
&= - E_{\boldsymbol{z}\sim p_{z}} \log (1-\boldsymbol{\hat{Y}}^1 [0]).
\end{split}
\end{equation}

The training of a semi-supervised GAN involves two alternating steps of stochastic gradient descent (SGD) or similar optimization methods in each iteration. First, the gradients of a discriminator $-\nabla_{\theta_D} L_{SEMI}$ are used to update the parameters $\theta_D$ of a discriminator for learning discriminative spectral-spatial HSI features. Second, the gradients of generators $-\nabla_{\theta_D} L_{G}$ are employed to update the parameters $\theta_G$ of a generator for improving the adversarial training of the semi-supervised GAN. 

\begin{figure}[!t]
\centering
\includegraphics[width=3in]{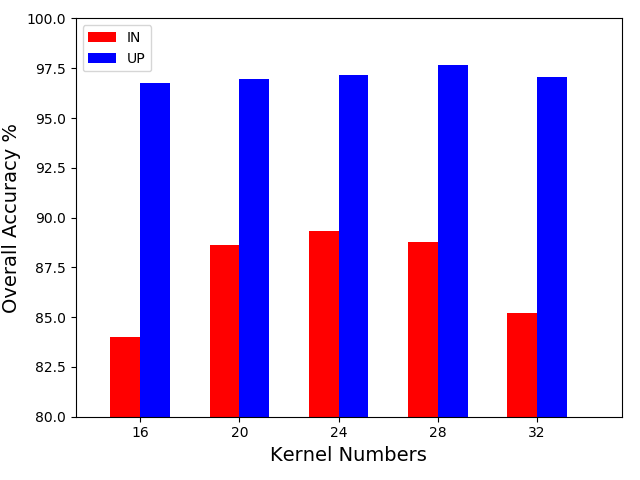}
\caption{Overall accuracies of semi-supervised GANs with different kernel numbers in their convolutiolnal and transposed convolutional layers using 300 labeled HSI samples for training.}
\label{fig_width}
\end{figure}

\begin{figure}[!t]
\centering
\includegraphics[width=3in]{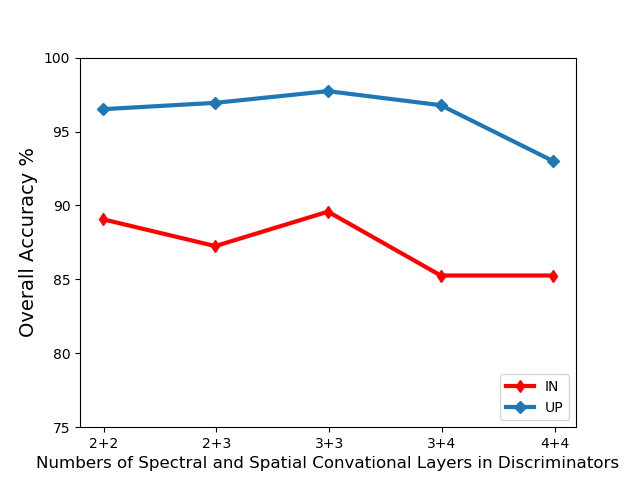}
\caption{
Overall accuracies of semi-supervised GANs that contain varying depths of spectral and spatial convolutional layers in their discriminators using 300 labeled HSI samples for training . The $x+y$ formation in the horizontal axis denotes a discriminator with $x$ spectral and $y$ spatial convolutional layers.
}
\label{fig_depth}
\end{figure}

\begin{table}[!t]
\renewcommand{\arraystretch}{1}

\centering
\captionsetup{justification=centering}

\caption{Overall Accuracies (\%) of semi-supervised GANs Using Different Numbers of Unlabeled and 200 Labeled HSI Samples in the IN and UP Datasets}
\label{table_1}
\begin{tabular}{ C{1.2cm} | C{1.6cm} | C{1cm} C{1cm} C{1cm}}
\hline
Datasets & Models & $0$ & $1000$ & $5000$ \\
\hline

IN & SPC-GAN\newline SPA-GAN\newline SS-GAN & 
$\boldsymbol{63.21}$ \newline $\boldsymbol{73.48}$ \newline $81.12$ &
$62.12$ \newline $71.28$ \newline $\boldsymbol{82.0}$ & 
$58.96$ \newline $67.62$ \newline $78.0$\\
\hline

UP & SPC-GAN\newline SPA-GAN\newline SS-GAN & 
$84.24$ \newline $91.01$ \newline $\boldsymbol{96.96}$ &
$\boldsymbol{84.69}$ \newline $\boldsymbol{91.74}$ \newline $95.76$ & 
$79.17$ \newline $87.35$ \newline $93.90$\\
\hline

\end{tabular}
\end{table}

\begin{table*}[!t]
\renewcommand{\arraystretch}{0.95}

\centering
\captionsetup{justification=centering}
\caption{Classification Results, Training, and Testing Times of Different Deep Learning Models Using 300 HSI Samples for the IN Dataset}
\label{table_2}
\centering
\begin{tabular}{C{1.2cm} |  C{1.2cm} | C{1.6cm} C{1.6cm} C{1.7cm} C{1.6cm} C{1.6cm} C{1.6cm} C{1.6cm}}
\hline
Class & Samples & 1D-GAN & AE-GAN & CNN-GAN & SS-CNN & SPC-GAN & SPA-GAN & SS-GAN \\
\hline

\hline
$1$& 3   	& $50.00$ &$0$		&$46.94$	 &$83.33$	&$66.67$	&$\boldsymbol{100.0}$	&$96.43$\\
$2$& 41  	& $51.98$ &$51.20$	&$46.45$	 &$77.88$	&$52.71$	&$64.48$	&$\boldsymbol{87.29}$\\
$3$& 29  	& $52.41$ &$38.75$	&$43.17$	 &$\boldsymbol{81.48}$	&$48.55$	&$61.49$	&$77.84$\\
$4$& 7   	& $35.38$ &$22.37$	&$47.66$	 &$76.47$	&$56.45$	&$81.56$	&$\boldsymbol{92.35}$\\
$5$& 14  	& $68.83$ &$49.74$	&$47.67$	 &$78.81$	&$69.44$	&$82.96$	&$\boldsymbol{92.64}$\\
$6$& 20  	& $87.30$ &$81.09$	&$63.37$	 &$87.14$	&$86.40$	&$93.98$	&$\boldsymbol{95.05}$\\
$7$& 2   	& $45.83$ &$0$		&$20.75$	 &$42.85$	&$67.86$	&$\boldsymbol{82.35}$	&$76.47$\\
$8$& 15  	& $86.86$ &$87.84$	&$79.13$	 &$89.45$	&$91.72$	&$90.75$	&$\boldsymbol{98.70}$\\
$9$& 3   	& $33.33$ &$0$		&$34.62$	 &$\boldsymbol{100.0}$	&$42.86$	&$45.45$	&$57.89$\\
$10$& 36 	& $39.29$ &$51.15$	&$61.37$	 &$77.94$	&$59.30$	&$78.83$	&$\boldsymbol{90.11}$\\
$11$& 64 	& $54.20$ &$64.83$	&$67.49$	 &$80.97$	&$72.96$	&$81.60$	&$\boldsymbol{95.19}$\\
$12$& 22 	& $45.57$ &$33.00$	&$34.20$	 &$62.52$	&$42.82$	&$53.68$	&$\boldsymbol{85.74}$\\
$13$& 4  	& $63.75$ &$81.31$	&$69.41$	 &$\boldsymbol{97.50}$	&$93.71$	&$87.32$	&$93.30$\\
$14$& 28 	& $80.36$ &$74.63$	&$77.32$	 &$88.63$	&$79.80$	&$82.32$	&$\boldsymbol{92.59}$\\
$15$& 10 	& $39.24$ &$47.91$	&$64.09$	 &$76.92$	&$66.76$	&$70.72$	&$\boldsymbol{78.74}$\\
$16$& $2$   & $98.63$ &$0$		&$84.29$ 	 &$\boldsymbol{100.0}$	&$77.78$	&$94.44$	&$95.29$\\
\hline
\multicolumn{2}{c}{OA (\%)} &$59.44$ & $60.26$ & $60.68$ & $81.07$ & $67.92$ & $76.65$ & $\boldsymbol{90.28}$\\

\multicolumn{2}{c}{AA (\%)} &$58.31$ & $42.74$ & $55.93$ & $81.37$ & $67.23$ & $78.23$ & $\boldsymbol{87.85}$\\

\multicolumn{2}{c}{$\kappa\times100$}  &$52.06$ & $54.24$ & $55.03$ & $78.21$ & $63.25$ & $73.30$ & $\boldsymbol{88.92}$\\
\hline

\multicolumn{2}{c}{Training (s)}  & $153.85$ & $217.70$ & $64.87$ & $139.55$ & $932.23$ & $233.32$ & $803.23$ \\

\multicolumn{2}{c}{Testing (s)}  &$0.59$ & $0.60$ & $0.35$ & $4.117$ & $5.88$ & $1.28$ & $5.09$ \\
\hline
\end{tabular}
\end{table*}

\begin{table*}[!t]
\renewcommand{\arraystretch}{0.95}

\centering
\captionsetup{justification=centering}
\caption{Classification Results, Training, and Testing Times of Different Deep Learning Models Using 300 HSI Samples for the UP Dataset}
\label{table_3}
\centering
\begin{tabular}{C{1.2cm} |  C{1.2cm} | C{1.6cm} C{1.6cm} C{1.7cm} C{1.6cm} C{1.6cm} C{1.6cm} C{1.6cm}}
\hline
Class & Samples & 1D-GAN & AE-GAN & CNN-GAN & SS-CNN & SPC-GAN & SPA-GAN & SS-GAN \\
\hline

\hline
$1$& 47  & $84.74$&$62.51$&$73.38$&$\boldsymbol{96.07}$&$84.74$&$91.10$&$95.62$\\
$2$& 132 & $92.50$&$92.02$&$90.17$&$97.57$&$87.31$&$96.93$&$\boldsymbol{99.49}$\\
$3$& 15  & $75.75$&$39.25$&$58.09$&$72.82$&$60.77$&$78.84$&$\boldsymbol{89.02}$\\
$4$& 20  & $93.46$&$84.55$&$98.39$&$\boldsymbol{99.37}$&$97.07$&$98.94$&$98.65$\\
$5$& 11  & $99.55$&$94.72$&$99.41$&$98.97$&$95.06$&$99.55$&$\boldsymbol{100.0}$\\
$6$& 35  & $86.77$&$62.72$&$74.21$&$98.18$&$86.70$&$92.71$&$\boldsymbol{99.09}$\\
$7$& 13  & $82.43$&$40.46$&$89.29$&$96.38$&$85.86$&$95.76$&$\boldsymbol{97.10}$\\
$8$& 21  & $73.79$&$51.78$&$83.65$&$82.81$&$75.85$&$86.88$&$\boldsymbol{92.54}$\\
$9$& 6   & $98.13$&$66.14$&$99.30$&$99.36$&$96.56$&$99.79$&$\boldsymbol{100.0}$\\
\hline
\multicolumn{2}{c}{OA (\%)} &$88.36$ & $75.10$ & $84.23$ & $95.04$ & $85.78$ & $93.97$ & $\boldsymbol{97.61}$\\

\multicolumn{2}{c}{AA (\%)} &$87.46$ & $66.02$ & $85.10$ & $93.50$ & $85.55$ & $93.39$ & $\boldsymbol{96.84}$\\

\multicolumn{2}{c}{$\kappa\times100$}  &$84.41$ & $67.07$ & $78.79$ & $93.40$ & $80.69$ & $91.98$ & $\boldsymbol{96.82}$\\
\hline

\multicolumn{2}{c}{ Training (s)}  & $107.27$ & $145.11$ & $64.71$ & $93.45$ & $647.68$ & $159.37$ & $527.46$ \\

\multicolumn{2}{c}{ Testing (s)}  &$2.06$ & $1.34$ & $1.76$ & $14.30$ & $18.38$ & $4.03$ & $15.36$ \\
\hline
\end{tabular}
\end{table*}

\begin{figure*}[!t]
\centering
\includegraphics[width=6in]{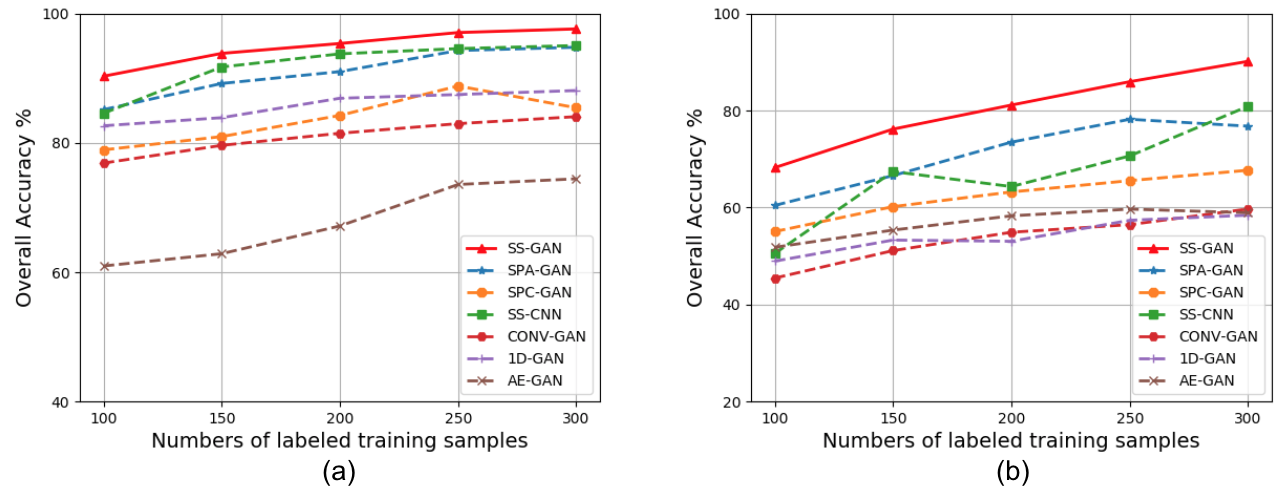}%
\caption{Overall accuracies of different semi-supervised GANs and the supervised benchmark SS-CNNs using from 100 to 300 HSI samples for training. (a) IN dataset. (b) UP dataset.}
\label{fig_train_res}
\end{figure*}

\subsection{GAN-CRF Model}

CRFs have been widely used to post-process image segmentation results because they can exploit the predictions of large numbers of unlabeled pixels to enhance image interpretation performance \cite{zheng2015conditional,cao2018hyperspectral}. Once a semi-supervised GAN has been built, we establish a conditional random field by using the softmax predictions of the trained semi-supervised GAN about unlabeled HSI cuboids to initialize random variables $\boldsymbol{Y}=\{\boldsymbol{y}\}$ that are conditioned on observed raw HSI pixels $\boldsymbol{X}$. According to Equation \eqref{crf}, the conditional probability distribution of this CRF takes the form:

\begin{equation}
Prob(\boldsymbol{y}|\boldsymbol{X}) = \frac{\exp(-E(\boldsymbol{y}|\boldsymbol{X}))}{\sum_{\boldsymbol{y}}\exp(-E(\boldsymbol{y}|\boldsymbol{X}))}.
\label{graph}
\end{equation}       

As illustrated in Figure 1, given that high correlations exists between HSI pixels $\{\boldsymbol{x_i}\}$ in both short- and long-range, we adopt a dense CRF \cite{Chen_Deeplab} that includes all pairwise connections between HSI pixels in the pairwise term of energy function to filter salt and pepper noises in homogeneous areas. The energy function of the dense CRF can be formulated as:

\begin{equation}
\label{energy}
E(\boldsymbol{Y} | \boldsymbol{X}) = U(\boldsymbol{Y}, \boldsymbol{X}) + P(\boldsymbol{Y},  \boldsymbol{X}),
\end{equation}
where $U(\cdot)$ and $P(\cdot)$ are the unary and pairwise terms of the energy function that is used to build the dense CRF. Specifically, the unary term represents the information cost of pixel-wise softmax predictions $\{\boldsymbol{y_i}\}$ and the binary term penalizes the wrong labeling of pixel pairs $\{\boldsymbol{x_i},\boldsymbol{x_j}\}$ with similar spectral signatures. These two terms are formulated as follows: 

\begin{equation}
U(\boldsymbol{Y}, \boldsymbol{X}) = \sum_{i} U(\boldsymbol{y_i} , \boldsymbol{X_i}) = \sum_{i} D(X_i; \theta_D),
\end{equation}

\begin{equation}
\begin{split}
P(\boldsymbol{Y}, \boldsymbol{X}) 
&= \sum_{i,j} P(\boldsymbol{y_i},\boldsymbol{y_j},\boldsymbol{x_i} , \boldsymbol{x_j}) \\
&= \sum_{i,j} \mu{(\boldsymbol{y_i},\boldsymbol{y_j})} K(\boldsymbol{x_i,x_j,l_i,l_j}),
\end{split}
\end{equation}
where $\boldsymbol{l_i}$ and $\boldsymbol{l_j}$ denote the locations of $\boldsymbol{x_i}$ and $\boldsymbol{x_j}$, respectively. $\mu(\cdot)$ is a compatibility function, and $K(\cdot)$ is a bilateral Gaussian kernel function. These two functions take the forms:

\begin{equation}
\label{compact}
\mu{(\boldsymbol{y_i},\boldsymbol{y_j})} = 
\begin{cases}
 	c,				& \text{if } \eta(\boldsymbol{y_i})\neq \eta(\boldsymbol{y_j})\\
    0,              & \text{otherwise}
\end{cases}
\end{equation}

\begin{equation}
\label{kernel}
K(\boldsymbol{x_i,x_j,l_i,l_j}) = \exp(-\frac{( \boldsymbol{{l_i-l_j}} )^{2}}{2{\theta}^{2}_{\alpha}} - \frac{( \boldsymbol{{x_i-x_j}} )^{2}}{2{\theta}^{2}_{\beta}}),
\end{equation}
where $\eta(\cdot)$ denotes a one-hot function. $\theta_\alpha$ and $\theta_\beta$ are two standard deviations of the bilateral Gaussian function. $c$ is a constant value that could be manually set. Random variables $\boldsymbol{Y} = \{\boldsymbol{y_i}\}$ of the established dense CRF is initialized to the softmax predictions of the trained discriminators $D(\boldsymbol{X}^2; \theta_D)$ of the semi-supervised GAN according to Equation \eqref{disc}. 

In a GAN-CRF model, a GAN is utilized to produce softax predictions about unlabeled HSI samples $\boldsymbol{X}^2$, and the post-processing CRF is independent of the GAN. Specifically, the predictions about a large numbers of unlabeled samples are used to initialize the unary term of the energy function that builds a dense CRF, and therefore the GAN-CRF model is more suitable in the case where only limited labeled samples are available. Because the energy function in Equation \eqref{energy} is an intractable problem, a function $Q(\boldsymbol{Y}|\boldsymbol{X})$ adopted to approximate the conditional probability distribution $Prob(\boldsymbol{Y}|\boldsymbol{X})$ of the CRF takes the form:

\begin{equation}
Q(\boldsymbol{Y}|\boldsymbol{X}) = \prod_{i} Q(\boldsymbol{y_i}|\boldsymbol{X}) \approx Prob(\boldsymbol{Y}|\boldsymbol{X}),
\end{equation}
in which the tractable function $Q(\boldsymbol{Y}|\boldsymbol{X})$ is close to $Prob(\boldsymbol{Y}|\boldsymbol{X})$ in terms of KL-distribution divergence. Then, the mean field approximation \cite{krahenbuhl2011efficient} is used to find an optimal solution of random variables ${\hat{\boldsymbol{Y}}}$ for the established dense CRF.

\section{Result and Discussion}

In this section, we introduce two challenging HSI datasets, set hyper-parameters of semi-supervised GANs, and evaluate GAN-CRF models and their competitors using performance metrics including the classification accuracy of each land cover class, overall accuracy (OA), average accuracy (AA), and kappa coefficient ($\kappa$). Additionally, we record training and testing times of all semi-supervised GANs to quantitatively assess their computational complexity. 

\subsection{Experimental Datasets}

Two most challenging and commonly studied HSI datasets -- the Indian Pines (IN) and the University of Pavia (UP) -- are used to evaluate the various types of semi-supervised GANs and GAN-CRF models for hyperspectral image classification. In both datasets, we randomly selected $\{100, 150, 200, 250, 300\}$ HSI cuboids with their annotations for training, and used the remaining cuboids for testing.  

As shown in Figure \ref{fig_in_res} (a) - (b), the IN dataset contains 16 vegetation classes and has $145 \times 145$ pixels with a spatial resolution of 20 m by pixel. 200 hyperspectral bands are used for this study and they range from 400 nm to 2500 nm. As illustrated in Figure \ref{fig_up_res} (a) - (b), the UP dataset includes 9 urban land cover types and has $610 \times 340$ pixels with a spatial resolution of 1.3 m by pixel. 103 hyperspectral bands are used for this research and they range from 430 nm to 860 nm. The numbers of labeled HSI samples for each land cover class for the IN and UP datasets can be found in Figure \ref{fig_in_res} and \ref{fig_up_res}, respectively. Given their relatively small numbers, the labeled hyperspectral groups ${\boldsymbol{X}^1}$ used for training contain at least two samples for each land cover class to avoid the situation that no sampled HSI cuboids are sampled for rare classes, especially in the IN dataset.

\subsection{Semi-supervised GAN Setting}

Figure \ref{fig_3} takes the UP dataset as an example to show the discriminator and generator of a semi-supervised GAN for HSI classification. In this semi-supervised GAN, the generator takes a $1\times1\times200$ vector of Gaussian noise as the input and outputs a $9\times9\times103$ fake HSI cuboid aiming to make the discriminator classify it as real data. Concurrently, a real $9\times9\times103$ HSI cuboids is randomly sampled from a raw HSI as the input of the discriminator. In this study, according to the result of a grid search, we set the learning rate to $0.0007$, batch size to $50$, and the spatial size of sampled HSI cuboids to $9\times9$. To avoid model collapse, we used Monte Carlo sampling \cite{saatci2017bayesian} to marginalize noise during training. Additionally, we adopted the Adam optimizer \cite{kingma2014adam} to alternatingly train the discriminator and generator. After the hyper-parameters of semi-supervised GANs are configured, we analyzed three factors that influence the classification performance of semi-supervised GANs.

First, the kernel number of convolutional and transposed convolutional layers affects the feature extraction and representation capacity of semi-supervised GANs. As illustrated in Figure \ref{fig_3}, the discriminator and generator of a semi-supervised GAN have the same kernel number in its convolutional and transposed convolutional layers. We tested different kernel numbers from 16 to 32 in an interval of 4 for all convolutional or transposed convolutional layers of semi-supervised GANs. As shown in Figure \ref{fig_width}, the semi-supervised GANs with 24 kernels in each layer achieved the highest classification accuracy using the IN dataset, and their counterparts with 28 kernels obtained the best classification performance using the UP dataset. These results are acquired in the 3000-epoch training for both datasets using randomly sampled 300 HSI cuboids. 

Second, the depth of the spectral-spatial discriminators in semi-supervised GANs also impacts their classification performance. Therefore, we assessed semi-supervised GANs with from 4 to 8 layers, which includes spectral and spatial convolutional layers, with the same hyper-parameter setting for each dataset. To make a fair comparison, we kept the generators of semi-supervised GANs have the same architecture as the generator in Figure \ref{fig_3}. As demonstrated in Figure \ref{fig_depth}, the semi-supervised GANs with 3 spectral and 3 spatial convolutional layers obtained the highest overall accuracies in both datasets. The fact that classification performance of semi-supervised GANs decreases with more convolutional layers than the optimal '$3+3$' architecture shows discriminators with deeper layers overfit the small number of labeled real HSI samples.

Third, to evaluate the influence of unlabeled real HSI cuboids, we tested three types of semi-supervised GANs using different numbers of unlabeled HSI samples for the IN and UP datasets. The three semi-supervised GANs are
the spectral GAN (SPC-GAN), and the spatial GAN (SPA-GAN), and the spectral-spatial GAN (SS-GAN). As shown in Figure \ref{fig_3}, the SS-GAN has both spectral and spatial learning blocks in its discriminator, and the SPC-GAN and SPA-GAN contain only spectral and spatial blocks, respectively. Again, we used the same setting of generators for all semi-supervised GANs as the generator in Figure \ref{fig_3}. Table \ref{table_1} shows that adding real unlabeled HSI samples for training contributes little to and adding more unlabeled samples even jeopardizes the semi-supervised HSI classification accuracy, which is caused by the different data distribution between labeled and unlabeled HSI samples. 



\subsection{Experimental Results}

We compared the proposed semi-supervised GANs to state-of-the-art GAN-based models, such as 1D-GAN \cite{zhan2018semisupervised} , AE-GAN\cite{Chen2014Deep}, and CNN-GAN\cite{zhu2018generative}. To demonstrate the effectiveness of the spectral-spatial architecture, we also compared spectral-spatial GANs (SS-GANs) that comprise three spectral and three spatial convolutional layers with their variants: SPC-GANs (three spectral layers) and SPA-GANs (three spatial layers). As shown in Figure {\ref{fig_3}}, we recorded the HSI classification results of the spectral-spatial convolutional neural networks (SS-CNNs) as important benchmarks. We kept the generators of all GANs the same, which consist of three spectral and four spatial transposed convolutions layers, each of which has 28 kernels. Then, we trained 3000 epochs for all GAN-based models, and set the input HSI cuboids with the same spatial size of $9\times 9\times$ for all methods that use spatial convolutional layers, and tuned the competitors to their optimal settings.   

Table \ref{table_2} and Table \ref{table_3} report the classification performance, including accuracy of all land cover classes, OAs, AAs, and Kappa coefficients, of the IN and UP datasets, respectively. In most cases, the proposed semi-supervised GANs perform better than the state-of-the-art GAN-based models. Interestingly, the supervised benchmark SS-CNNs perform slightly better than SPA-GANs, which shows the discriminative feature learning capacity of spectral and spatial convolutional layers. More importantly, the SS-GANs achieved the highest overall classification accuracies ($90.28\%$ and $97.61\%$ OAs for the IN and UP datasets, respectively) among all GAN-based models and the SS-CNNs. It is worth noting that he semi-supervised SS-GANs outperform fully supervised SS-CNNs in IN and UP datasets with $9.21\%$ and $2.57\%$, respectively, which shows that the generated samples are helpful for improving classification accuracy. These results demonstrate the effectiveness of spectral-spatial convolutional architectures and semi-supervised adversarial training. Additionally, Table \ref{table_2} and Table \ref{table_3} also show the training and testing times of all models, which indicate the computational costs of these models. All experiments were conducted using an NVIDIA TITAN Xp graphical processing unit (GPU). In both datasets, the SPC-GANs are the slowest to train and the SS-GANs take about 6 times longer for training than SS-CNNs.  

To test the robustness of the SS-GANs and their competitors, we randomly sampled different numbers of labeled HSI cuboids in an interval of 50 from 100 to 300 to train these semi-supervised GANs and SS-CNNs for the IN and UP datasets. As shown in Figure \ref{fig_train_res}, the classification performance of SPA-GANs is comparable to that of SS-CNNs. AE-GANs perform clearly worse than other models because their fully connected layers fail to take the spectral-spatial characteristics  of HSI samples into account. More importantly, the proposed SS-GANs consistently outperform their semi-supervised competitors and SS-CNNs in both datasets. These results demonstrate the importance of accounting for the attributes of training data to design deep learning models, which is in line with the report of \cite{zhong2017spectral}.

\begin{figure*}[!t]
\centering
\includegraphics[width=7.2in]{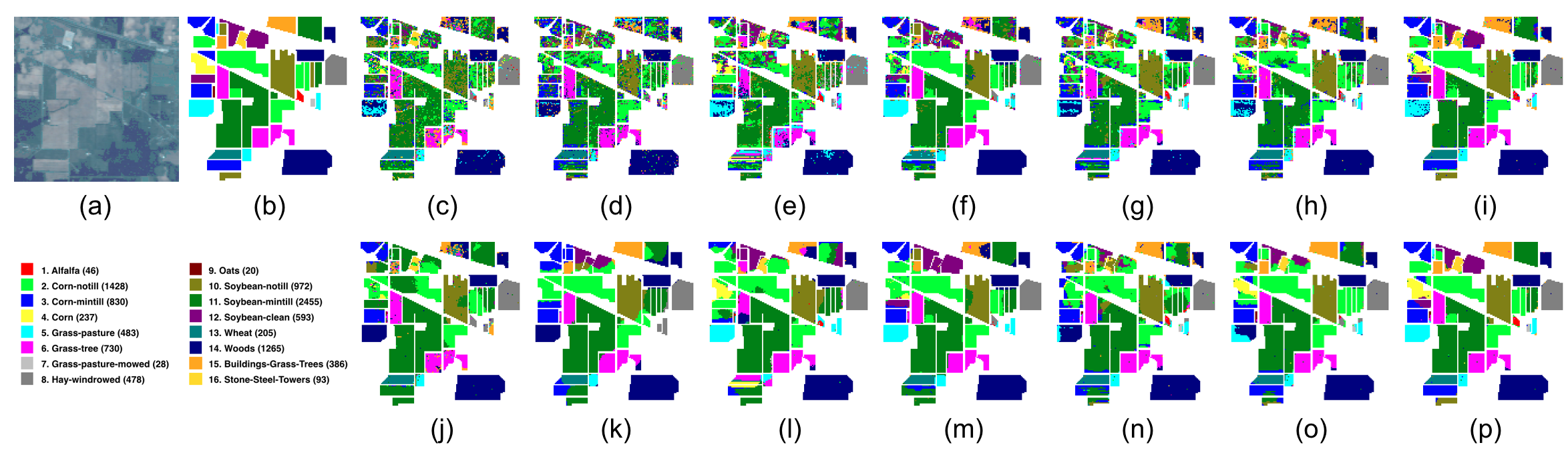}%
\caption{Classification results of semi-supervised GAN models, a supervised CNN, and their refined counterparts by adding dense CRFs using 300 labeled HSI samples for the IN dataset. (a) False color image. (b) Ground truth labels. (c) - (i) Classification maps of 1D-GAN, AE-GAN, CNN-GAN, SS-CNN, SPC-GAN, SPC-GAN, and SS-GAN. (j) - (p) Classification maps of 1D-GAN-CRF, AE-GAN-CRF, CNN-GAN-CRF, SS-CNN-CRF, SPC-GAN-CRF, SPA-GAN-CRF, and SS-GAN-CRF.}
\label{fig_in_res}
\end{figure*}

\begin{figure*}[!t]
\centering\
\includegraphics[width=7.2in]{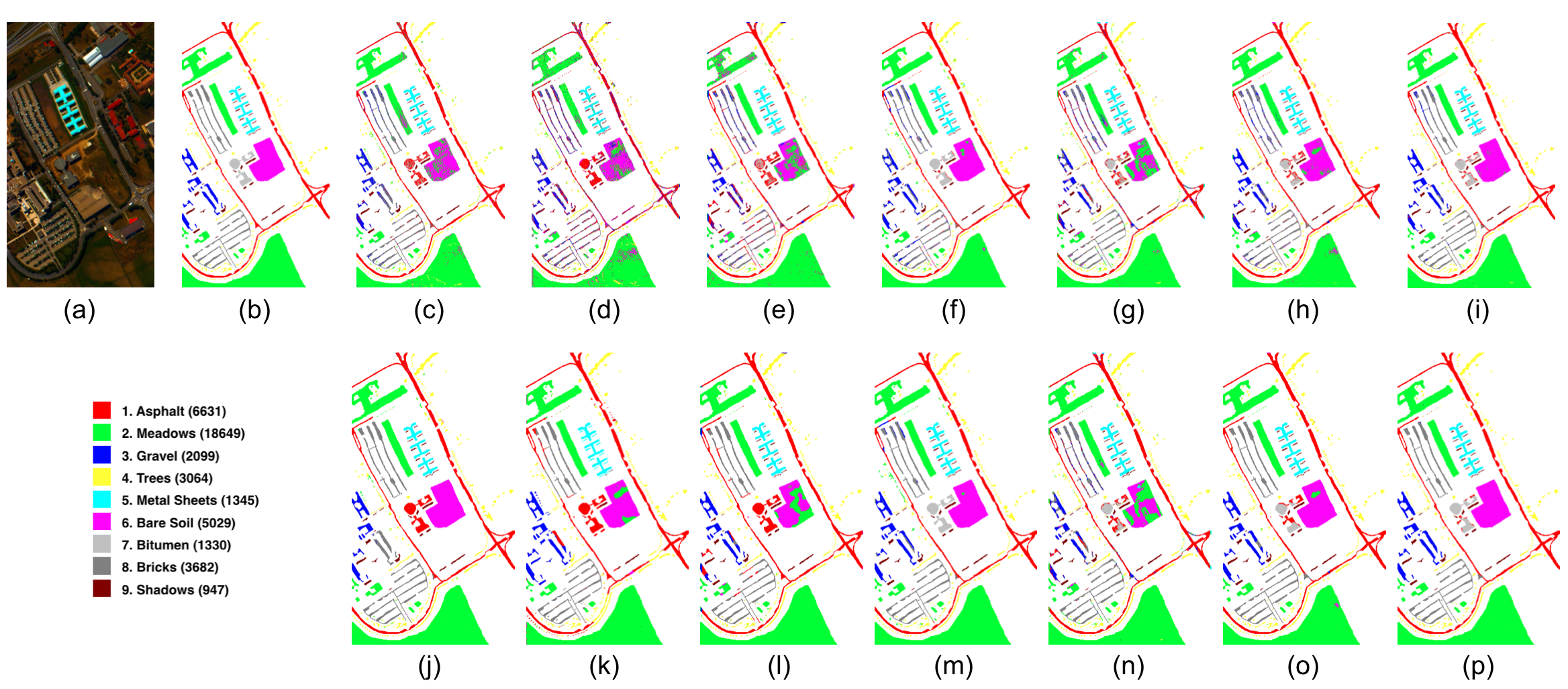}%
\caption{Classification results of semi-supervised GAN models, a supervised CNN, and their refined counterparts by adding dense CRFs using 300 labeled HSI samples for the UP dataset. (a) False color image. (b) Ground truth labels. (c) - (i) Classification maps of 1D-GAN, AE-GAN, CNN-GAN, SS-CNN, SPC-GAN, SPC-GAN, and SS-GAN. (j) - (p) Classification maps of 1D-GAN-CRF, AE-GAN-CRF, CNN-GAN-CRF, SS-CNN-CRF, SPC-GAN-CRF, SPA-GAN-CRF, and SS-GAN-CRF.}
\label{fig_up_res}
\end{figure*}

In this study, we used the three most prominent principal component analysis (PCA) channels of HSI $\boldsymbol{X}$ instead of raw HSI cuboids to facilitate the mean field approximation of the dense CRF. As shown in Table \ref{table_crf}, SS-GANs and SS-GAN-CRF models perform better than their competitors, and GAN-CRF models significantly enhance the classification performance of those models without integrating dense CRFs,  Moreover, Figure \ref{fig_in_res} and Figure \ref{fig_up_res} show the classification maps of all semi-supervised GANs and all GAN-CRF models. The qualitative results of these classification maps are in line with the quantitative report of Table \ref{table_crf}. The SS-GAN-CRF models deliver the most accurate overall classification accuracies($96.30\%$ and $99.31\%$ OAs for the IN and UP datasets, respectively) and smoothest classification maps for both HSI datasets, because the SS-GANs learn the most discriminative spectral-spatial features and dense CRFs consider long-range correlations between similar HSI samples. Therefore, these classification outcomes validate the feasibility of integrating semi-supervised deep learning and graph models given limited labeled HSI samples for training.

\begin{table}[!t]
\renewcommand{\arraystretch}{0.95}
\caption{Overall Accuracies (\%) of deep learning models and their refined results by adding dense CRFs using 300 labeled HSI samples for training}
\label{table_crf}
\centering
\begin{tabular}{c | c | c | c | c}
\hline
{} & \multicolumn{2}{c}{IN Dataset} & \multicolumn{2}{c}{UP Dataset}\\
\hline
Models & w/o CRF & w/ CRF & w/o CRF& w/ CRF\\
\hline
1D-GAN& $59.44$ & $70.41$ & $88.36$ & $94.41$ \\
AE-GAN& $60.26$ & $76.08$ & $75.10$ & $90.44$ \\
CNN-GAN& $60.28$ & $73.83$ & $84.23$ & $90.42$ \\
SS-CNN& $81.07$ & $87.66$ & $95.04$ & $98.05$ \\
SPC-GAN& $68.92$ & $74.64$ & $85.78$ & $88.13$ \\
SPA-GAN& $76.65$ & $85.64$ & $93.97$ & $97.57$ \\
SS-GAN& $\boldsymbol{90.28}$ & $\boldsymbol{96.30}$ & $\boldsymbol{97.61}$ & $\boldsymbol{99.31}$ \\
\hline
\end{tabular}
\end{table}

\subsection{Discussion}

There are three differences between the GAN-CRF framework and the original GAN proposed in [11]. First,  GAN-CRF models take the spectral-spatial characteristics of HSI data into account for both the discriminators and generators. Second, the discriminators in the semi-supervised framework extend the softmax predictions $\hat{\boldsymbol{y}}$ of a GAN from two classes (fake/real) to $1+n_y$ classes, where $n_y$ represents the number of land cover classes. Third, a post-processing dense CRF has been built on conditional random variables that are initialized to the softmax outputs of the trained GANs to filter salt and pepper noises in homogenous areas.

The GAN-CRF models incorporate the CRF as a post-processing step and build a graph upon the learned features and the softmax outputs of discriminators to refine HSI classification maps. Compared with those CRFs adopted in previous articles \cite{zhong2010learning,zhao2015detail}, the fully connected CRFs consider the long-range correlations between HSI samples. This property helps GAN-CRF models to better filter noises in the homogeneous areas of some land cover classes. Compared to just a supervised discriminator, a GAN-CRF model integrates the advantages of deep learning models and probabilistic graph models and improves HSI classification accuracy. There are two main reasons for this improvement: 1) the synthetic HSI samples produced by generators help discriminators to learn more robust and discriminative features; 2) the subsequent dense CRFs consider the spectral similarity and spatial closeness of HSI samples to refine the softmax outputs conditional on these samples using the trained discriminators of GANs.

We gain four major insights from the semi-supervised HSI classification outcomes of GANs and GAN-CRF models in both datasets. First, by taking the characteristics of training data into account, the discriminators of SS-GANs extract discriminative HSI features and achieve better classification accuracy. 
Second, generators of SS-GANs learn feature representation by producing synthetic HSI samples, and in turn make discriminators more robust to adversaries and learn more discriminative features. Therefore, this adversarial training enables semi-supervised GANs to deliver superior classification outcomes to supervised deep learning models. Third, adding unlabeled real HSI samples to train semi-supervised GANs marginally improves or even jeopardizes the HSI classification results. Fourth, dense CRFs take the classification maps generated by semi-supervised GANs as an initialization and smooth the noisy classification maps by adding a pairwise term that imposes the correlation between similar or neighboring pixels from input HSIs.

\section{Conclusion}

In this paper, we have proposed a semi-supervised GAN-CRF framework to address three commonly occurring challenges for HSI classification: the high spectral dimensionality of training data, the small numbers of labeled samples, and the noisy classification maps generated by deep learning models. First, we designed four consecutively structured convolutional and transposed convolutional layers to take the spectral-spatial characteristics of HSIs into consideration. Second, we established semi-supervised GANs, each of which comprises a generator and a discriminator, to extract discriminative features and to learn feature representation of HSI samples. Third, we integrated a probabilistic graphical model with a semi-supervised deep learning model to refine HSI classification maps. The experimental results using two of the most widely studied and challenging HSI datasets demonstrate that the spectral-spatial GANs (SS-GANs) perform the best among all semi-supervised GAN-based models and supervised benchmark models, and subsequently that the spectral-spatial GAN-CRF (SS-GAN-CRF) models achieved state-of-the-art performance for semi-supervised HSI classification.

The GAN-CRF models demonstrate an effective way to integrate two mainstream pixel-wise HSI classification methods --- deep learning and probabilistic graphical models --- and this framework can be easily generalized to other image interpretation cases. These two models have complementary advantages in the sense that deep learning models focus on discriminative feature extraction and implicit feature representation, and graph models emphasize the smoothness prior of images that is crucial for accurate classification and segmentation. However, the GAN-CRF framework presents a two-step setting because the dense CRFs function as a post-processing step to refine the classification maps generated by GANs. 

The contributions of this work mainly focus on validating the feasibility to integrate these two parts and show a way to implement this target. Therefore, a joint training framework is our future task, and current models need to be redesigned to achieve this goal. For example, we could make the discriminator of GAN a local semantic segmentation network and change the generator accordingly. The reason of the separated training lies in the different roles of Semi-supervised GAN and fully-connected CRF. The GAN aims for training a discriminative model in a semi-supervised way and then using the trained model to generate pixelwise conditional probabilities. In contrast, the adoption of CRF considers the pixelwise classification prediction holistically and adds structural constraints on top of it.  Therefore, we will continue this research line for imposing graph constraints on the convolutional layers of deep learning models to construct an end-to-end trainable framework.


%



\ifCLASSOPTIONcaptionsoff
  \newpage
\fi



%

\bibliographystyle{IEEEtran}
\bibliography{bibfile}

\begin{thebibliography}{10}
\providecommand{\url}[1]{#1}
\csname url@samestyle\endcsname
\providecommand{\newblock}{\relax}
\providecommand{\bibinfo}[2]{#2}
\providecommand{\BIBentrySTDinterwordspacing}{\spaceskip=0pt\relax}
\providecommand{\BIBentryALTinterwordstretchfactor}{4}
\providecommand{\BIBentryALTinterwordspacing}{\spaceskip=\fontdimen2\font plus
\BIBentryALTinterwordstretchfactor\fontdimen3\font minus
  \fontdimen4\font\relax}
\providecommand{\BIBforeignlanguage}[2]{{%
\expandafter\ifx\csname l@#1\endcsname\relax
\typeout{** WARNING: IEEEtran.bst: No hyphenation pattern has been}%
\typeout{** loaded for the language `#1'. Using the pattern for}%
\typeout{** the default language instead.}%
\else
\language=\csname l@#1\endcsname
\fi
#2}}
\providecommand{\BIBdecl}{\relax}
\BIBdecl

\bibitem{li2014hyperspectral}
H.~Li, G.~Xiao, T.~Xia, Y.~Y. Tang, and L.~Li, ``Hyperspectral image
  classification using functional data analysis,'' \emph{IEEE Trans. Cybern.},
  vol.~44, no.~9, pp. 1544--1555, 2014.

\bibitem{jia20183}
S.~Jia, L.~Shen, J.~Zhu, and Q.~Li, ``A 3-d gabor phase-based coding and
  matching framework for hyperspectral imagery classification,'' \emph{IEEE
  Trans. Cybern.}, vol.~48, no.~4, pp. 1176--1188, 2018.

\bibitem{yuan2016hyperspectral}
Y.~Yuan, J.~Lin, and Q.~Wang, ``Hyperspectral image classification via
  multitask joint sparse representation and stepwise mrf optimization,''
  \emph{IEEE Trans. Cybern.}, vol.~46, no.~12, pp. 2966--2977, 2016.

\bibitem{zhou2016learning}
Y.~Zhou and Y.~Wei, ``Learning hierarchical spectral--spatial features for
  hyperspectral image classification,'' \emph{IEEE Trans. Cybern.}, vol.~46,
  no.~7, pp. 1667--1678, 2016.

\bibitem{Chen2016Deep}
Y.~Chen, H.~Jiang, C.~Li, X.~Jia, and P.~Ghamisi, ``Deep feature extraction and
  classification of hyperspectral images based on convolutional neural
  networks,'' \emph{IEEE Trans. Geosci. Remote Sens.}, vol.~54, no.~10, pp.
  6232--6251, 2016.

\bibitem{zhong2017spectral}
Z.~Zhong, J.~Li, Z.~Luo, and M.~Chapman, ``Spectral-spatial residual network
  for hyperspectral image classification: A 3-d deep learning framework,''
  \emph{IEEE Trans. Geosci. Remote Sens.}, vol.~56, no.~2, pp. 847--858, 2018.

\bibitem{Luo2018}
F.~Luo, B.~Du, L.~Zhang, L.~Zhang, and D.~Tao, ``Feature learning using
  spectral-spatial hypergraph discriminant analysis for hyperspectral image,''
  \emph{IEEE Trans. Cybern.}, vol.~49, no.~7, pp. 2406--2419, 2019.

\bibitem{tarabalka2010segmentation}
Y.~Tarabalka, J.~Chanussot, and J.~A. Benediktsson, ``Segmentation and
  classification of hyperspectral images using minimum spanning forest grown
  from automatically selected markers,'' \emph{IEEE Trans. Syst., Man, Cybern.
  B, Cybern.}, vol.~40, no.~5, pp. 1267--1279, 2010.

\bibitem{yuan2017spectral}
H.~Yuan and Y.~Y. Tang, ``Spectral--spatial shared linear regression for
  hyperspectral image classification,'' \emph{IEEE Trans. Cybern.}, vol.~47,
  no.~4, pp. 934--945, 2017.

\bibitem{Zhao2016Spec}
W.~Zhao and S.~Du, ``Spectral–spatial feature extraction for hyperspectral
  image classification: A dimension reduction and deep learning approach,''
  \emph{IEEE Trans. Geosci. Remote Sens.}, vol.~54, no.~8, pp. 4544--4554,
  2016.

\bibitem{Li2017Hyperspectral}
W.~Li, G.~Wu, F.~Zhang, and Q.~Du, ``Hyperspectral image classification using
  deep pixel-pair features,'' \emph{IEEE Trans. Geosci. Remote Sens.}, vol.~55,
  no.~2, pp. 844--853, 2017.

\bibitem{Chen2014Deep}
Y.~Chen, Z.~Lin, X.~Zhao, G.~Wang, and Y.~Gu, ``Deep learning-based
  classification of hyperspectral data,'' \emph{IEEE J. Sel. Topics Appl. Earth
  Observ. Remote Sens.}, vol.~7, no.~6, pp. 2094--2107, 2014.

\bibitem{Chen2015Spec}
Y.~Chen, X.~Zhao, and X.~Jia, ``Spectral-spatial classification of
  hyperspectral data based on deep belief network,'' \emph{IEEE J. Sel. Topics
  Appl. Earth Observ. Remote Sens.}, vol.~8, no.~6, pp. 2381--2392, 2015.

\bibitem{Krizhevsky2012Imagenet}
A.~Krizhevsky, I.~Sutskever, and G.~E. Hinton, ``Imagenet classification with
  deep convolutional neural networks,'' in \emph{in Proc. Adv. Neural Inf.
  Process. Syst.}, Conference Proceedings, pp. 1106--1114.

\bibitem{LeCun2015Deep}
Y.~LeCun, Y.~Bengio, and G.~Hinton, ``Deep learning,'' \emph{Nature}, vol. 521,
  no. 7553, pp. 436--444, 2015.

\bibitem{Mnih2015Human}
V.~Mnih, K.~Kavukcuoglu, D.~Silver, A.~A. Rusu, J.~Veness, M.~G. Bellemare,
  A.~Graves, M.~Riedmiller, A.~K. Fidjeland, and G.~Ostrovski, ``Human-level
  control through deep reinforcement learning,'' \emph{Nature}, vol. 518, no.
  7540, pp. 529--533, 2015.

\bibitem{cao2018hyperspectral}
X.~Cao, F.~Zhou, L.~Xu, D.~Meng, Z.~Xu, and J.~Paisley, ``Hyperspectral image
  classification with markov random fields and a convolutional neural
  network,'' \emph{IEEE Trans. Image Process.}, vol.~27, no.~5, pp. 2354--2367,
  2018.

\bibitem{Tarabalka2010SVM}
Y.~Tarabalka, M.~Fauvel, J.~Chanussot, and J.~A. Benediktsson, ``Svm- and
  mrf-based method for accurate classification of hyperspectral images,''
  \emph{IEEE Trans. Geosci. Remote Sens. Lett.}, vol.~7, no.~4, pp. 736--740,
  2010.

\bibitem{zhong2014hybrid}
Y.~Zhong, J.~Zhao, and L.~Zhang, ``A hybrid object-oriented conditional random
  field classification framework for high spatial resolution remote sensing
  imagery,'' \emph{IEEE Trans. Geosci. Remote Sens.}, vol.~52, no.~11, pp.
  7023--7037, 2014.

\bibitem{yang2014semi}
L.~Yang, S.~Yang, P.~Jin, and R.~Zhang, ``Semi-supervised hyperspectral image
  classification using spatio-spectral laplacian support vector machine,''
  \emph{IEEE Trans. Geosci. Remote Sens. Lett.}, vol.~11, no.~3, pp. 651--655,
  2014.

\bibitem{zhong2011modeling}
P.~Zhong and R.~Wang, ``Modeling and classifying hyperspectral imagery by crfs
  with sparse higher order potentials,'' \emph{IEEE Trans. Geosci. Remote
  Sens.}, vol.~49, no.~2, pp. 688--705, 2011.

\bibitem{ji2014spectral}
R.~Ji, Y.~Gao, R.~Hong, Q.~Liu, D.~Tao, and X.~Li, ``Spectral-spatial
  constraint hyperspectral image classification,'' \emph{IEEE Transactions on
  Geoscience and Remote Sensing}, vol.~52, no.~3, pp. 1811--1824, 2014.

\bibitem{zhan2018semisupervised}
Y.~Zhan, D.~Hu, Y.~Wang, and X.~Yu, ``Semisupervised hyperspectral image
  classification based on generative adversarial networks,'' \emph{IEEE Trans.
  Geosci. Remote Sens. Lett.}, vol.~15, no.~2, pp. 212--216, 2018.

\bibitem{zhu2018generative}
L.~Zhu, Y.~Chen, P.~Ghamisi, and J.~A. Benediktsson, ``Generative adversarial
  networks for hyperspectral image classification,'' \emph{IEEE Trans. Geosci.
  Remote Sens.}, 2018.

\bibitem{Goodfellow}
I.~Goodfellow, J.~Pouget-Abadie, M.~Mirza, B.~Xu, D.~Warde-Farley, S.~Ozair,
  A.~Courville, and Y.~Bengio, ``Generative adversarial nets,'' in \emph{in
  Proc. Adv. Neural Inf. Process. Syst.}, 2014, pp. 2672--2680.

\bibitem{Chen_Deeplab}
L.-C. Chen, G.~Papandreou, I.~Kokkinos, K.~Murphy, and A.~L. Yuille, ``Deeplab:
  Semantic image segmentation with deep convolutional nets, atrous convolution,
  and fully connected crfs,'' \emph{IEEE Trans. Pattern Anal. Mach. Intell.},
  vol.~40, no.~4, pp. 834--848, 2018.

\bibitem{Salimans}
T.~Salimans, I.~Goodfellow, W.~Zaremba, V.~Cheung, A.~Radford, and X.~Chen,
  ``Improved techniques for training gans,'' in \emph{in Proc. Adv. Neural Inf.
  Process. Syst.}, 2016, pp. 2234--2242.

\bibitem{Radford}
A.~Radford, L.~Metz, and S.~Chintala, ``Unsupervised representation learning
  with deep convolutional generative adversarial networks,'' \emph{arXiv
  preprint arXiv:1511.06434}, 2015.

\bibitem{saatci2017bayesian}
Y.~Saatci and A.~G. Wilson, ``Bayesian gan,'' in \emph{in Proc. Adv. Neural
  Inf. Process. Syst.}, 2017, pp. 3622--3631.

\bibitem{lecun1998gradient}
Y.~LeCun, L.~Bottou, Y.~Bengio, and P.~Haffner, ``Gradient-based learning
  applied to document recognition,'' \emph{Proc. IEEE}, vol.~86, no.~11, pp.
  2278--2324, 1998.

\bibitem{yang2004two}
J.~Yang, D.~Zhang, A.~F. Frangi, and J.-y. Yang, ``Two-dimensional pca: a new
  approach to appearance-based face representation and recognition,''
  \emph{Trans. Pattern Anal. Mach. Intell.}, vol.~26, no.~1, pp. 131--137,
  2004.

\bibitem{zhao2016high}
J.~Zhao, Y.~Zhong, H.~Shu, and L.~Zhang, ``High-resolution image classification
  integrating spectral-spatial-location cues by conditional random fields,''
  \emph{IEEE Trans. Image Process.}, vol.~25, no.~9, pp. 4033--4045, 2016.

\bibitem{krahenbuhl2011efficient}
P.~Kr{\"a}henb{\"u}hl and V.~Koltun, ``Efficient inference in fully connected
  crfs with gaussian edge potentials,'' in \emph{in Proc. Adv. Neural Inf.
  Process. Syst.}, 2011, pp. 109--117.

\bibitem{Zhang2014Hyperspectral}
L.~Zhang, L.~Zhang, D.~Tao, X.~Huang, and B.~Du, ``Hyperspectral remote sensing
  image subpixel target detection based on supervised metric learning,''
  \emph{IEEE Trans. Geosci. Remote Sens.}, vol.~52, no.~8, pp. 4955--4965,
  2014.

\bibitem{Ioffe2015Batch}
S.~Ioffe and C.~Szegedy, ``Batch normalization: Accelerating deep network
  training by reducing internal covariate shift,'' in \emph{in Proc. 32nd Int.
  Conf. on Mach. Learn.}, pp. 448--456.

\bibitem{dai2017good}
Z.~Dai, Z.~Yang, F.~Yang, W.~W. Cohen, and R.~R. Salakhutdinov, ``Good
  semi-supervised learning that requires a bad gan,'' in \emph{in Proc. Adv.
  Neural Inf. Process. Syst.}, 2017, pp. 6513--6523.

\bibitem{zheng2015conditional}
S.~Zheng, S.~Jayasumana, B.~Romera-Paredes, V.~Vineet, Z.~Su, D.~Du, C.~Huang,
  and P.~H. Torr, ``Conditional random fields as recurrent neural networks,''
  in \emph{Proc. IEEE Conf. Comput. Vis. Pattern Recognit.}, 2015, pp.
  1529--1537.

\bibitem{kingma2014adam}
D.~P. Kingma and J.~Ba, ``Adam: A method for stochastic optimization,''
  \emph{arXiv preprint arXiv:1412.6980}, 2014.

\bibitem{zhong2010learning}
P.~Zhong and R.~Wang, ``Learning conditional random fields for classification
  of hyperspectral images,'' \emph{IEEE Trans. Image Process.}, vol.~19, no.~7,
  pp. 1890--1907, 2010.

\bibitem{zhao2015detail}
J.~Zhao, Y.~Zhong, and L.~Zhang, ``Detail-preserving smoothing classifier based
  on conditional random fields for high spatial resolution remote sensing
  imagery,'' \emph{IEEE Trans. Geosci. Remote Sens.}, vol.~53, no.~5, pp.
  2440--2452, 2015.

\end{thebibliography}




%










 \end{document}